\documentstyle[12pt,aaspp4,psfig]{article}

\def\ML{$M/L$}
\def\cent {$\omega$\thinspace Centauri}

\def\kms{km s$^{-1}$}

\def\CaII{Ca\,{\sc ii}}
\def\gtsim{\ {\raise-0.5ex\hbox{$\buildrel>\over\sim$}}\ }
\def\ltsim{\ {\raise-0.5ex\hbox{$\buildrel<\over\sim$}}\ }
\slugcomment{To appear in The Astronomical Journal, January 1998 issue}

\begin{document}

\title{Radial Velocities of Globular Clusters in the Giant Elliptical
Galaxy NGC~1399} 

\author {Dante Minniti$^{1,2}$, Markus Kissler-Patig$^{3,2,4}$, Paul
Goudfrooij$^{5,6,2}$, and Georges Meylan$^2$}

\altaffiltext{1}{Lawrence Livermore National Laboratory, Livermore, CA
94550, USA\\ E-mail: dminniti@llnl.gov}

\altaffiltext{2}{European Southern Observatory,
Karl-Schwarzschild-Str. 2, D-85748 Garching bei M\"unchen, Germany\\
E-mail: gmeylan@eso.org} 

\altaffiltext{3}{Lick Observatory, University of California, Santa
Cruz, CA 95064, USA\\ E-mail: mkissler@ucolick.org}

\altaffiltext{4}{Sternwarte der Universit\"at Bonn, Auf dem H\"ugel
71, D-53121 Bonn, Germany}

\altaffiltext{5}{Space Telescope Science Institute, 3700 San Martin
Drive, Baltimore, MD 21218, USA\\ 
E-mail: goudfroo@stsci.edu}

\altaffiltext{6}{Affiliated with the Astrophysics Division, Space
Science Department, European Space Agency, ESTEC, Postbus 299, NL-2200
AG Noordwijk, The Netherlands}

\begin{abstract}
We report on the radial velocity measurements of 18 globular clusters
and one dE galaxy
in the field of the giant elliptical galaxy NGC~1399, the dominant
galaxy of the Fornax cluster.  We present also the radial velocity
measurements of some candidates for young and super metal-rich
globular clusters, which turn
out to be foreground or background objects, viz. 28 stars and 7
galaxies.  The bonafide globular clusters are selected, on the basis of their
magnitudes and colors, to be metal-rich.  For this cluster
sample we measure a mean radial velocity of $v_{rad} = 1353 \pm 79$ \kms,
and a velocity dispersion of $\sigma = 338 \pm 56$ \kms.  
Using a few different mass estimators, this implies \ML\ values in
the range $50-130$ within a radius of 28 kpc, consistent with \ML\
rising with radius.  Our velocity dispersion estimate $\sigma$ is intermediate
between the value computed from the integrated stellar light at smaller
radii, and that computed at large radii from recent X-ray
observations.
\end{abstract}


\section{Introduction}
A globular cluster contains about $10^5$ stars of different masses,
generally with similar ages and metallicities. From both
dynamical and stellar population points of view, globular clusters
appear significantly simpler than galaxies.  Because they can be
observed to much larger distances than individual stars, properties of
globular clusters can be used to study the formation and evolution of
their host galaxies.  In particular, the discovery of a very large
number of clusters associated with the giant elliptical galaxy M87
(Baum 1955; Racine 1968; Forte et al.\ 1981; Strom et al.\ 1981)
generated a number of fascinating questions about the formation of
cluster systems of giant elliptical galaxies (e.g., van den Bergh
1990; Ashman \& Zepf 1992; Harris 1995; Kissler-Patig 1997; Forbes et
al.\ 1997a).

Particularly interesting is the discovery of bimodality in the color
distribution of the globular cluster systems of some elliptical
galaxies, which has been generally interpreted as a bimodal metallicity
distribution (e.g., Zepf \& Ashman 1993; Geisler et al.\ 1996
Whitmore et al.\ 1996).  This suggests that the globular cluster
systems of some ellipticals may be a superposition of different
cluster populations.

More progress in the understanding of the formation of cluster systems
in galaxies is coming from large radial velocity surveys (e.g.,
Mould et al.\ 1987, 1990; Grillmair et al.\ 1994; Hui et al.\ 1995;
Bridges et al.\ 1997, Cohen \& Ryzhov 1997). 
Globular cluster systems are ideal probes for
studying the dynamics of the outer parts of galaxies, where the
surface brightness of the galaxy itself becomes too faint to provide
useful kinematic data. Furthermore, the masses of globular clusters
are negligible with respect to the mass of the galaxy so that they
directly trace the underlying gravitational potential.  Radial
velocity surveys of globular cluster systems thus provide a powerful
means to constrain the dynamical predictions of elliptical galaxy
formation scenarios (e.g., isotropy, dark matter content).  This is
nicely illustrated by the recent work by Cohen \& Ryzhov (1997), who
are able for the first time to accurately map the radial dependence of
\ML\ in M87, using radial velocities from {\it Keck\/} spectroscopy of a
large sample of 205 globular clusters.

In this paper we concentrate on the globular cluster system of
NGC~1399, the dominant E1\,{\it pec\/} galaxy (de Vaucouleurs et al.\
1991) in the center of the Fornax cluster.  The `peculiarity' of this
galaxy is that it is a cD galaxy with an unusually
extended stellar envelope.  NGC~1399 is surrounded by a very rich
population of globular clusters (Hanes \& Harris 1986; Geisler \&
Forte 1990; Wagner et al.\ 1991; Bridges et al.\ 1991). The clusters
have a bimodal color distribution, as confirmed independently by
different groups (Ostrov, Geisler \& Forte 1993; Kissler-Patig et al.\
1997). Based on the photometry, these authors suggest two or more
enrichment or formation episodes.

The only previous velocity measurements for globular clusters in
NGC~1399 are those by Grillmair et al.\ (1994), who found a large
velocity dispersion ($\sigma = 388 \pm\ 54$ \kms). This velocity
dispersion is larger than that of the Fornax cluster itself, which has
$\sigma = 307-325$ \kms\ (Ferguson \& Sandage 1990). The mean radial
velocity of the clusters studied by Grillmair et al.\ is $v_{rad} = 1518
\pm 91$ \kms, which is actually larger than that of NGC~1399 itself.  They
inferred a large \ML\ ratio for this galaxy, viz. $M/L = 79 \pm 20$.

Figure~1a displays the radial velocity distribution of the 47 globular
clusters in NGC~1399 as measured by Grillmair et al.\ (1994).  We
notice that this velocity distribution is non-gaussian, and likely
bimodal.  The heliocentric radial velocity 
$v_{rad} = 1443$ \kms\ (Tully 1988) 
of the dominant galaxy NGC~1399 is close to the velocity of one of the peaks,
while the other peak is close to the velocity of the nearby galaxy
NGC~1404 with $v_{rad} = 1908$ \kms, located at about 10 arcmin
($\sim$\,47 kpc) SE of the center of NGC~1399.
For comparison, the velocities listed in the RC3 (de Vaucouleurs et
al.\ 1991) for these galaxies are for NGC~1399:
$v_{rad}(GSR) = 1323$ \kms, $v_{rad}(opt) = 1447$ \kms\ 
and for NGC~1404: $v_{rad}(GSR) = 1805$ \kms,  $v_{rad}(opt) = 1929$ \kms.  

Figure~1b displays the radial velocity distribution of 67
galaxy members of the Fornax cluster, as measured by Ferguson (1989).
We also noticed that the velocity distribution of these galaxies
appears non-gaussian.  The radial velocity of NGC~1399 seems to be
offset from the mean velocity of the Fornax cluster, being smaller by
about $100-200$ \kms, depending on the adopted velocity of this
galaxy: e.g., Ferguson (1989) takes $v_{rad} = 1430$ \kms.
This offset is not surprising, given that the
centroid of the X-ray emission of the Fornax cluster is also offset
from NGC~1399 (Ikebe et al.\ 1997). It is well known that some central
galaxies are displaced from the center of their corresponding cluster
potential well (cf.\ Hill et al.\ 1988; Beers et al.\ 1991; Oegerle \&
Hill 1994), and this is usually taken as an indication of substructure
within the clusters (see the analysis by Bird 1994).

Given these indications of a complex system, 
the goal of this paper is two-fold: 

1.\ to measure the radial velocities of a selected sample of
globular clusters around
NGC~1399 and to investigate the kinematics of the globular cluster
system;

2.\ to derive, from the velocity dispersion of these globular clusters,
the \ML\ for this galaxy at a radius intermediate between those probed
by the galaxy integrated light and by the X-ray-emitting gas.

\section{Selection of the Sample}

Defining a sample of globular clusters that traces the galaxy rather
than the galaxy cluster potential is important, since some globulars
may have been stripped away from galaxies and populate the intergalactic
medium (Muzzio 1987; Bassino et al.\ 1994; West et al.\ 1994). 
For example, Theuns \& Warren 
(1997) showed evidence for intergalactic stars in the Fornax cluster,
based on the discovery of 10 planetary nebulae.
In relation to this, the X-ray emission mapped by ASCA has been
interpreted as the superposition of concentrated emission from
NGC~1399 itself, and more extended emission from the whole galaxy
cluster (Ikebe et al.\ 1996). 

There are observational indications that the kinematics of a globular
cluster system may well depend on metallicity.  For example, in the
specific case of the merger NGC~5128, Hui et al.\ (1995) combined the
NGC~5128 globular cluster velocities from Harris et al. (1988) and
Sharples (1988) with 
the photometry of Harris et al.  (1992). Dividing up the
sample at [Fe/H]~=~$-1$ into a metal-poor and a metal-rich sample,
they showed that the metal-rich globular 
cluster system of NGC~5128 rotates more rapidly and has smaller
velocity dispersion than the metal-poor one.  This is consistent with
the fact that the metal-rich globular clusters are more centrally
concentrated (Minniti et al.\ 1996).

Indeed, metal-rich clusters appear to be better tracers of the
underlying metal-rich stellar populations, in elliptical galaxies
(e.g., Forbes et al.\ 1997a) as well as in the Milky Way bulge (e.g.,
Minniti 1995). In order to study the kinematics of this population, we
select metal-rich clusters in NGC~1399 on the basis of the
colors measured by Kissler-Patig et al.\ (1997). 
The color distribution of NGC~1399 globular clusters is
bimodal (Ostrov, Geisler \& Forte 1993; Kissler-Patig et al.\
1997). The color distribution shown by Kissler-Patig et al.\ (1997)
ranges from about $V\!-\!I = 0.5$ to 1.5, with peaks at $V\!-\!I =
0.9$ and $1.2$, the second peak corresponding to the mean color of the
integrated light of NGC 1399 ($V\!-\!I = 1.24$, Poulain 1988).  It is
worth mentioning that the absolute location of these peaks in terms of
[Fe/H] is uncertain due to the age-metallicity degeneracy of optical
colors (see, e.g., Worthey 1994, Vazdekis et al.\ 1996). However, the two 
strong
peaks differ by more than $\Delta$\,[Fe/H] = 0.5 dex in the mean if
interpreted as a pure metallicity difference.

We selected globular cluster candidates over the whole color range
mentioned above (see Figure 2), focusing on metal-rich globular
clusters which should have stronger spectral lines, which is an
advantage for measuring radial velocities.  These red candidates have
$1.0 < (V\!-\!I)_0 < 1.5$, where we have adopted $E_{V-I} = 0.02$ for this
field (Kohle et al.\ 1996).  These colors are equivalent to selecting
clusters with [Fe/H] \gtsim $-1.0$ (see Kissler-Patig et al.  1997,
their Figure 4), or a population similar to the metal-rich globular
clusters associated with the Milky Way bulge (Minniti 1995). 

We included a few candidates in the range $0.7 < (V\!-\!I)_0 < 1.0$, but
all turned out to be stars, or to have too weak metal lines to measure
good velocities. We further included even bluer and slightly brighter
objects, in order to check whether NGC 1399 is hosting young- or
intermediate-age globular clusters as seen in recent mergers
(e.g.,~NGC~7252, [Whitmore et al.\ 1993], NGC~1275, [Holtzman et
al. 1992; N{\o}rgaard--Nielsen et al.\ 1993], NGC~5128, [Minniti et
al.\ 1996] or NGC~5018 [Hilker \& Kissler-Patig 1996]). All of our
selected best candidates had good enough 
spectra to identify them as clear foreground stars and exclude the
hypothesis of younger globular clusters. 

Finally, we included very red
objects with $(V\!-\!I)_0 > 1.5$, usually discarded in color
distributions as background objects, in order to confirm their
non-globular cluster nature. Here too the test was successful in the
sense that very red objects indeed turned out to be either M stars or
background galaxies, and neither very metal--rich, nor reddened
globular clusters.
Note that the errors in our photometry are small ($\sigma_{V-I} <
0.04$ mag at $18.5<V<21$), which is essential for the selection of
clusters belonging to the metal-rich tail of the color distribution,
and to the other groups considered.

As to the magnitude limit, we adopt for NGC~1399 a distance modulus of
$m - M = 31.0$ after Kohle et al.\ (1996).  At this distance, 1 arcmin
corresponds to 4.6 kpc.  This distance modulus is intermediate between
the distance moduli of $30.6$ and $31.5$, adopted by Grillmair et al.\
(1994) and Ikebe et al.\ (1996), respectively.  Therefore, in order to
minimize foreground objects but to be complete at the bright end of
the globular cluster luminosity function, we chose objects with
$V>18.5$ (corresponding to $M_V<-12.5$), except for the very blue
objects that were assumed to be potential young globular clusters, and
therefore brighter, for which we chose candidates with $16.5<V<18.5$
(i.e.~$-12.5<M_V<-14.5$).  Figure 2 shows the color-magnitude diagram
of 18 globular clusters with available spectra, 
along with 26 foreground stars. The color distribution of objects in the
field of NGC~1399 from Kissler-Patig et al. (1997) is shown for comparison.

These tests indicate that globular cluster sample selection based on 
accurate magnitude and color limits can be very successful in deriving a
clean globular cluster sample.

\section{Observations}

Spectroscopy of NGC 1399 clusters was carried out during four
half-nights, in late 1995 and early 1996, with the ESO 3.5-m New
Technology Telescope (NTT) equipped with the ESO Multi-Mode Instrument
(EMMI) in Multi-Object Spectroscopy (MOS) mode.

The detector was a thin, back-illuminated CCD of type Tektronix
TK2048EB Grade 2, having 2048$\times$2048 pixels (of size 24 $\mu$m =
0\farcs268 pixel$^{-1}$). We used grism No.\ 4 (2.8 \AA\ pixel$^{-1}$,
blazed at 6500 \AA). The wavelength range is $5500 - 10000$ \AA,
although the actual wavelength coverage of the individual spectra
varied according to the location of the slitlets in the MOS masks.
The slitlets had a width of 1\farcs34, resulting in a spectral
resolution of 7.5~\AA.

We decided to take spectra in the spectral range including the \CaII\
triplet ($\lambda\lambda$\,8498,\,8542,\,8662 \AA) since the \CaII\
index is an excellent metallicity indicator, being independent of age
for objects older than about 1 Gyr (Bica \& Alloin 1987, Armandroff \&
Zinn 1988).  Unfortunately, the majority of our spectra do not have
high enough S/N ratios to allow accurate metallicity abundance
measurements.

Astrometry and photometry for the observed globular cluster candidates
in NGC~1399 is available from Kissler-Patig et al.\ (1997). However,
precise astrometry for MOS targets alone is not sufficient to produce
MOS plates, since an accurate map of the distortion in the focal plane
of the NTT was not available. To overcome this problem, short-exposure
($R$ band, exp.\ time = 5 min) ``acquisition'' images were kindly
taken in advance by J. van Loon, from which the ($x, y$) positions
of our targets were derived to produce the MOS masks.

The effective punching area for the MOS masks is 5$'$\,$\times$\,8$'$.
Three MOS masks were made to include the maximum possible number of
globular cluster candidates during the available time. In each mask,
two or three slitlets were punched at positions corresponding to
bright ``reference'' Galactic stars in the field of view, which were
used to check the precision of the telescope pointing.

Spectra were taken on October 7, 1995 for a field centered NE of NGC~1399 
(our field 2), with a total exposure time of 150 minutes. The seeing was
1\farcs0, as judged from the direct image immediately preceding the
MOS observation. Spectra for a field centered NW of NGC~1399 (our field 4) were
obtained on November 7, 1995, with a seeing of 0\farcs9 and a total
exposure time of 135 minutes, and the spectra for a field centered SE 
of NGC~1399 (our field 1) were obtained on January 30 and January 31,
1996, with an average seeing of 0\farcs9 and a total exposure time
of 130 + 110 minutes in the two half--nights.

Multiple exposures were acquired during each night. Individual
exposures were typically of 1800-s duration, and interleaved with
short imaging exposures to correct for small pointing errors due to
e.g., flexure and differential atmospheric refraction (the corrections
needed were only of order 0\farcs3 or smaller; the airmass was always
below 1.3).

The third mask, located in the SE of NGC~1399, has twice as many
coadded exposures (6) than the other two masks, located at the NW and
NE of the center of this galaxy. Therefore,  
we were able to extract more spectra in the SE field, and with higher S/N
ratios. On the other hand, the NW field had the lowest S/N and objects
fainter than V$=20.7$ could barely be measured. 
The SW field could not be observed due to lack of time.

The observations were made during bright time, with typical sky
surface brightness of $\mu_I = 19.2$ mag/arcsec$^2$. The underlying
``background'' brightness of the galaxy is more problematic only in
the inner regions of NGC~1399, where some of the targets are located:
E.g., $\mu_I = 18.5$ mag/arcsec$^2$ at 25 arcsec from the galaxy center
(Goudfrooij et al.\ 1994). These objects require a longer exposure
time, and the large background variations account for a wide range of
S/N ratios achieved in the different spectra.

\section{Data Reduction}

A total of 20 bias frames were taken in the afternoon of each
observing night. These were averaged together to form a ``superbias''
frame, in which the effect of read-out noise is rendered
negligible. The superbias frame was subtracted from each ``science''
frame, after having been properly scaled to match the bias level in
the serial overscan region of each observation.  Multiple dome flat
field exposures were obtained for each MOS mask with the grism in
place, using an external lamp. After combining the flat field images
for each individual mask, the flat fields were normalized to an
average value of unity. This was done for each slitlet individually by
dividing the measured wavelength response of the flat field lamp by a
third-order polynomial fit.

The two-dimensional spectra of each star and cluster candidate were
extracted from the flatfielded MOS frames and wavelength calibrated
using the corresponding He-Ar calibration lamp exposure.  Residual
zeropoint errors were corrected using the night-sky lines which are
numerous and prominent in the wavelength region covered.  The
extracted spectra were combined and appeared largely free of cosmic
ray hits; the few residual affected pixels were interpolated over by
hand.  The sky background was derived for each target individually by
fitting a Chebyshev polynomial to the background regions in the
slitlets, along as well as perpendicular to the dispersion. The fitted
background spectrum was then subtracted from each individual spectrum.

The final spectra were cross-correlated with 2-3 stellar template
spectra using the method of Tonry \& Davis (1979), incorporated in the
package FXCOR within IRAF.  Night-sky
lines, which are very numerous and prominent in the wavelength region
covered, were used for the zero-point of the individual templates
observed simultaneously with the same setup.  
Although no radial velocity standards from the IAU lists
were observed, the zero point of our velocities is good to $50$ \kms.
Cross-correlation peaks with $R < 2.5$ (this is the R ratio of
Tonry and Davis 1979, defined as the relative height of the main
cross-correlation peak with respect to the neighboring peaks),
were discarded, as their
velocities are not reliable enough, judging from the cross
correlations against the different templates. A total of 6 objects did
not yield spectra with a S/N ratio high enough to meet this criterion.
The typical FWHM of the cross-correlation peaks is about 500 \kms, and
the abscissa of the peak itself is defined to better than 100 \kms.
Although the \CaII\ triplet used to measure radial velocities is
usually very strong in globular cluster spectra, it is weaker in the
metal-poor globulars.  Therefore, we have concentrated on the
metal-rich ones, for which velocities can be measured with higher
confidence in relatively shorter exposure times.

Table 1 lists the globular cluster data: column 1 contains the object ID,
columns 2, 3, 4 and 5 contain the x and y coordinates in arcsec from
the center of the galaxy, and the J1950 equatorial coordinates,
respectively, following Kissler-Patig et al. (1997);
columns 6  and 7
list their magnitudes and colors, respectively; column 8 lists the
heliocentric radial velocities in \kms; columns 10 and 11 list the
errors $\Delta_{Vr}$ and $\epsilon_{Vr}$, corresponding to the maximum
difference in velocities computed with respect to the various
templates, and these given by the cross correlations package,
respectively. Table 2 lists similar data for the background galaxies
and foreground stars in our sample. Their classifications are listed in
the last column.

We cross-checked the positions of our globular clusters with those of
Grillmair et al.\ (1994), finding two objects in  common, \#119 and
\#406. The velocity differences for these clusters are $\Delta V = 
57$ and $-63$ \kms, respectively, showing good consistency.
A couple of fainter cluster candidates that are included in the list
of Grillmair et al.\ (1994) had too low S/N to provide
accurate velocities.

\section{Foreground Stars in the Sample}

We obtained spectra for 28 stars (see Table 2), some of which (7) were
deliberately chosen as standard stars (to use as pointing check and as
radial velocity templates), and some (21) turned out to be failed
young cluster candidates and reddened or super metal-rich cluster candidates.
These spectra turn out to be very useful also to
estimate the uncertainties, as discussed below.

Figure 3 shows the dependence of the radial velocity on the $V$
magnitudes and $V\!-\!I$ colors for the foreground stars and the NGC~1399
clusters.  There is a clear segregation in radial velocity between
the foreground stars and the
globular clusters and background galaxies.  All objects with $v_{rad}
\la 400$ \kms\ are foreground stars and all objects brighter than
$V \sim 19.5$ are foreground stars or galaxies, but {\it not\/}
globular clusters.

The mean velocity of the foreground stars ($v_{rad} = 65 \pm 27$ \kms, $N
= 28$) matches the expected velocity of halo stars in this direction
of the sky ($v_{rad} = 220 \cos l \sin b = 110$ \kms).  Sommer-Larsen
(1987) predicts, in this direction of the sky, a mean velocity and
velocity dispersion of $v_{rad} = 110$ \kms\ and $\sigma = 115$ \kms,
respectively.  These values are confirmed -- to within the errors --
by a small sample of stars ($N=6$) measured in this field by Grillmair
et al.\ (1994), who give $v_{rad}=144 \pm 254$ \kms\ and $\sigma = 269
\pm 167$ \kms.

The intrinsic velocity error can also be estimated by comparing the
observed stellar velocity dispersion ($\sigma = 143 \pm 19$, $N=28$)
with that expected for the MW stars along this line-of-sight ($\sigma
= 115$, cf.\ Sommer-Larsen 1987).  Subtracting these in quadrature we
obtain an intrinsic error per measurement of $\sigma_0 = 85$ \kms,
which agrees with the typical errors listed in Table 1. This should be
considered an optimistic estimate of the errors, because the stars are brighter 
than the globular clusters in the mean.

We can also divide up the stellar sample according to their brightness
(see Figure 3). The 11 stars with $V < 18$ have $v_{rad} = 53 \pm 17$
\kms, and $\sigma = 56 \pm 12$ \kms, consistent with a mixture of
foreground disk and halo dwarf stars. The fainter stars with $18<V<21$
have $v_{rad} = 72 \pm 44$ \kms, and $\sigma = 180 \pm 31$ \kms,
consistent with a dominant halo population. This difference in 
velocity dispersion between the two subsamples persists even
if we take into account that the fainter stars have larger errors.
However, there are no detailed kinematic studies
published for such faint stars to compare with these results.

\section{Globular Cluster Results}

For the adopted distance modulus of NGC~1399, viz. $m - M = 31.0$
(Kohle et al.\ 1996), the brightest globular clusters members of
NGC~1399 have, in our sample, $V \sim 19.5$ i.e. $M_V = -11.5$.  This
is significantly brighter than the brightest globular clusters in the
Galaxy and M31, viz. \cent\ with $M_V = -10.07$, and Mayall~II = G1
with $M_V = -10.55$, respectively. However, more luminous globular
clusters with $M_V \leq -10.7$ are known in other well studied giant galaxies
from photometric (e.g., Whitmore et al.\ 1995; Elson et al.\ 1996),
and spectroscopic studies  (e.g. Mould et al. 1990).  Note that the object
at $V = 18.5$, $V\!-\!I = 1.48$ (our reddest ``globular cluster''),
which has $M_V = -12.5$ was identified as a compact dwarf galaxy on
the images after light profile analysis (Hilker priv.~com.). We will 
include this object in the subsequent analysis.

A relaxed virialized system will have a near-Maxwellian velocity
distribution, which implies that the observed radial velocity
distribution should be Gaussian to a first approximation. Figure 4
shows the velocity distribution for our sample of 18 globular clusters
in the field of NGC~1399. This velocity distribution is 
different in appearance from that of Grillmair et al.\ (1994,
Fig.~1). Given the small number of objects, however, the difference is not
statistically significant. 
Figure 4 shows one dominant peak at $v_{rad} \sim
1350$ \kms, which is coincident with the lowest peak of their
distribution.  Fitting a Gaussian  to this distribution
gives $v_{rad} = 1353 \pm 79$ \kms, and $\sigma = 338
\pm 56$ \kms\ for $N=18$ objects. We argue that this 
value for $\sigma$ may have no great physical meaning because {\it (i)\/}
the velocity distribution may not be Gaussian (as it shows
two peaks, Grillmair et al.\ 1994), 
and {\it (ii)\/} there could be some contamination by
globular clusters associated either with NGC~1404 or with the
intergalactic medium. 

The globular cluster system of NGC~1404 was studied by Richtler et
al.\ (1992), Hanes \& Harris (1986), Forbes et al.\ (1997b), and
Blakeslee \& Tonry (1996). In particular, Forbes et al.\ (1997b)
suggest that a large fraction of intermediate-metallicity clusters may
have been stripped by NGC~1399, in order to account for the low
specific frequency of globulars in NGC~1404.  
In fact, some of the clusters populating 
the high velocity tail of the distribution are indeed located in the
direction of NGC~1404.  Note that with a larger sample we should be
able to distinguish whether the possible bimodality is indeed due to
contamination from NGC~1404, or if this is due to all the clusters
being in nearly circular orbits.  The latter situation has been argued
to be appropriate for the galaxy M49 (NGC~4472), which also has a
bimodal velocity distribution with peaks separated by about 500 \kms\
(Mould et al.\ 1990).

If we eliminate the cluster with highest radial velocity, arguing that it
is a potential member of NGC~1404 or the Fornax cluster of galaxies itself,
then $v_{rad} = 1309 \pm 71$ \kms, and $\sigma = 293
\pm 50$ \kms\ for $N=17$ objects.
Figure 5 shows the radial dependence of velocity dispersion for
different tracers in the field of NGC~1399.  These tracers include the
integrated stellar light (from Franx et al.\ 1989; Bicknell et al.\
1989; Winsall \& Freeman 1993), planetary nebulae (from Arnaboldi et
al.\ 1994), globular clusters (from Grillmair et al.\ 1994, and the
present paper), and galaxies belonging to the Fornax cluster (from
Ferguson 1989). Our values of $\sigma$, computed using N=17 and N=18 objects,
follow the trend of the integrated light from the galaxy.

Arnaboldi et al.\ (1994) have detected a rotation in the
system of planetary nebulae in the NGC~1399 halo. They measure a
rotation curve with 1.09 \kms\ arcsec$^{-1}$ and position angle
$-35^{\circ}$. Accepting this figure rotation does not change the
velocity dispersion of the present cluster system significantly,
although the mean velocity in field 1 is lower than that of field 4,
showing the opposite effect.  The present sample lacks objects in the
SW region, and is more extended in the NW-SE direction, as shown in
Figure 6.  More radial velocity measurements with adequate spatial
distribution are needed to derive a figure rotation for the metal-rich
cluster system of NGC~1399.  In addition, any fitting for rotation
would have to assess the probability of contamination from the
clusters associated with the nearby elliptical NGC~1404 and/or the
intergalactic medium.

We can use one of the Jeans equations of stellar dynamics (Binney \&
Tremaine 1987), in order to estimate the mass of the system:
$$M(r) = - {\sigma_r^2\, r \over G} \, (d\ln \rho/d\ln r + d\ln
\sigma_r^2/d\ln r + 2 \beta),$$
where $M(r)$ is the total mass inside the radius $r$,
$\sigma_r$ is the radial component of the velocity dispersion of the
test particles, $\rho$ is the spatial density of the test particles,
and $\beta$ = $ 1 - \sigma_{\theta}^2 / \sigma_r^2$ is a measure of the
degree of anisotropy of the velocity distribution at each point.
Given the lack of observational constraints on the radial gradient of
the velocity dispersion, we consider the simplifying hypotheses of
isothermality and isotropy, i.e. $\sigma_r$ = $\sigma$, $d\ln \rho/d\ln
r$ = 0, and $\beta$ = 0.

Another question concerns the spatial distribution of the clusters.
The spatial distribution of the NGC~1399 cluster system
has been studied by Ostrov, Geisler, \& Forte (1993) and
Kissler-Patig et al.\ (1997). The total surface density profile
follows a power law of with $n=-1.7$. The globular clusters in NGC
1399 follow a clearly flatter distribution then in nearby smaller
ellipticals; however, the distribution still matches that of the
galaxy light due to the extended  cD envelope of NGC 1399
(Kissler-Patig et al.\ 1997). No clear difference
between the spatial distribution of the blue and red globular cluster
populations have been found. 

Grillmair et al.\ (1994) used the Jeans equation,
adopting spherical symmetry and circular
orbits, in order to compute a lower limit of $M/L = 79\pm 20$ for the cluster
system of NGC~1399.  The velocity dispersion found here 
implies a slightly smaller \ML: since $M \propto
\sigma^2$, the $M/L$ value corresponding to $\sigma = 388$ \kms\
of Grillmair et al.\ (1994)
yields $M/L \sim 60$ for $\sigma = 338$
\kms. However, since $M/L \propto distance$, and since we have adopted
a distance modulus $m-M_0=31.0$ which is 0.4 mag larger than that of
Grillmair et al.\ (1994), this value would be reduced further to $M/L = 50$.  

We have used the projected mass estimator (Heisler et al.\ 1985), again
under the hypothesis of a spherical distribution of matter and
isotropy of the velocity dispersion (see Perelmuter et al.\ 1995
and Bridges et al.\ 1997 for similar applications of this method to other
galaxies).  The projected mass $M_P$ for an
isotropic velocity ellipsoid is:
$$M_P = { 32 \over \pi NG} \times \Sigma_1^{18} (v_i - v_o)^2 R_i,$$
where $v_i$ is the radial velocity of the considered globular cluster,
$v_o$ is either the radial velocity of the galaxy or the mean radial
velocity of the cluster sample, and $R_i$ the globular cluster
distance on the plane of the sky from the center of the galaxy.  For
the 18 globular clusters with a mean radial velocity of 1353 \kms\ 
for which $\sigma = 338\pm 56$, 
we obtain $M_P = 5.0 \times
10^{12}~M_{\odot}$.  This mass estimate, along with $L = 3.9\times
10^{10} ~L_{\odot}$ (Grillmair et al.\ 1994, scaled to our adopted
distance), gives $M_P/L \sim 129$. 
The error in this quantity is dominated by the small sample size and
the different underlying theoretical assumptions.  E.g., adopting
radial instead of isotropic orbits for the globular clusters increases
the mass estimate by a factor of two.

Other mass estimators (Heisler et al.\ 1985) applied to the same
sample give the following results: the virial mass $M_{VT} = 4.2
\times 10^{12}~M_{\odot}$ implies $M_{VT}/L \sim 109$, the median mass
$M_M = 3.3 \times 10^{12}~M_{\odot}$ implies $M_M/L \sim 87$, and the
average mass $M_{AV} = 4.2 \times 10^{12}~M_{\odot}$ implies $M/L \sim
109$.  The virial, projected, and average masses share the same
sensitivity to interlopers.  For example, if there are interlopers from the
intracluster medium, or from the nearby galaxy NGC~1404, these would tend to 
have higher velocities.  Eliminating the object with the
largest radial velocity (\#407 with $V_r = 2107$ \kms) from the present sample of 18
globular clusters, the above $M/L$ values decrease to
$M_P/L \sim 81$ for projected mass, $M_{VT/L} \sim 78$ for virial
mass, $M_M/L \sim 68$ for median mass, and $M_{AV}/L \sim 67$ average
mass.  These \ML\ values, obtained from the globular cluster system of
NGC~1399, are similar to the corresponding values obtained with the same
mass estimators by Huchra \& Brodie (1987) from the globular cluster
system of M87.

Contrary to the velocity dispersion determinations, which suffer only
from the observational errors on the radial velocities, the
mass-to-luminosity estimates accumulate the errors associated with many
various observed parameters, such as the total luminosity within
a given radius, the distance modulus and the radial velocity of the
host galaxy, not to mention the different theoretical assumptions.
Considering the large uncertainties on the \ML\ determinations,
reliable comparisons with the earlier study by Grillmair et al.\
(1994) can be made only for the above determination using similar
assumptions, namely use of the Jeans equation and providing $M/L \sim 50$. 
This value is somewhat smaller than those reported in previous
studies at large distance from the center of NGC~1399, but it is in
accord with the \ML\ values computed from the X-ray emission (Ikebe et
al.\ 1996).  Note that this value of \ML\ is larger than that of a
typical old stellar population ($1< M/L <10$), supporting the
existence of dark matter. We also confirm that \ML\ increases with radius,
ruling out a constant \ML.  A similar result, although much more
accurate due to the large number of high quality {\it Keck\/} spectra
available, is obtained by Cohen \& Ryzhov (1997) for the cluster
system of M87.  The same conclusion can be reached by looking at
Figure 5, which shows that our velocity dispersion measurement is
intermediate between those from the outermost integrated
stellar light (from Franx et al.\ 1989; Bicknell et al.\ 1989; Winsall
\& Freeman 1993) and the estimate from velocities of galaxies
belonging to the Fornax cluster (from Ferguson 1989).

\section{Conclusions} 

We have measured velocities for 53 objects in the field of NGC~1399,
including 7 background galaxies, one dwarf galaxy in Fornax, 18
globular cluster candidates, and 28 foreground stars.  Located in the
central region of the Fornax cluster of galaxies, NGC~1399 is a
particularly interesting system to use as a reference, because it does
not seem to host young or intermediate age globular clusters as seen
in recent or past mergers such as NGC 7252, NGC 1275, NGC~5128 or NGC
5018.  The main conclusions are:

$\bullet$ We find that all candidates with $V<19.5$ are either
foreground stars or background galaxies, and {\it not} young globular clusters.  
We also find that the very red
($(V\!-\!I)_0 > 1.5$) objects in the color
distribution are {\it not} (very metal-rich or reddened) globular
clusters.

$\bullet$ The radial velocity distribution of metal rich globular
clusters is different from the distribution of the whole
population. In particular, 
it shows a dominant peak centered on the
radial velocity of NGC~1399 itself.

$\bullet$ Assuming that the metal-rich globular clusters are a better
tracer of the underlying galaxy than the metal-poor ones, we have a
sample that better traces the potential of this galaxy.  We derive a
velocity dispersion of $\sigma = 338\pm 56$ \kms\ 
(or $\sigma = 293\pm 50$ \kms\ eliminating the cluster with largest velocity). 
These values are somewhat smaller than that measured by Grillmair et al.\
(1994), who did not discriminate against more metal-poor globular clusters.

$\bullet$ The kinematics of the globular cluster system of NGC~1399
are complicated, which is perhaps related to the presence of different
populations as inferred from the bimodal color distribution, and
assumptions of dynamical equilibrium or isotropy of the whole system
are not warranted.  In particular, we note that NGC~1399 is offset
from the center of the Fornax cluster potential well, both in velocity
and in position in the sky. This suggests the presence of substructure
in the Fornax cluster, and in order to unravel the dynamics of the
system, velocities for more tracers are needed.

$\bullet$ Ignoring the previous conclusion, we derive, in a way
similar to Grillmair et al.\ (1994), an estimate of $M/L \sim 50$
at a mean distance of $r = 5.5$ arcmin. This value joins nicely the
measurements based on the integrated galaxy light in the inner regions
with those based on recent X-ray emission measurements in the outer
regions (Ikebe et al.\ 1996). This would argue for \ML\ increasing
with radius in this galaxy, though at more moderate rates than
previously thought.  Other mass estimators provide \ML\ in the range
50-130, although with large uncertainties.

The total
number of data points is too small to do a comprehensive statistical analysis,
as $N>50$ points are required (e.g. Bird 1994).
Consequently, larger radial velocity samples are needed. Until they
become available, we can neither precisely evaluate the 
rotation of the metal-rich globular cluster population, nor
investigate its dynamics.  Other questions that remain to be answered
are (i) the extent to which clusters that belong to NGC~1404
contribute to the measured dispersions, and (ii) whether or not some
clusters from NGC~1404 been stripped by NGC~1399 in the past, as
proposed by Forbes et al.\ (1997b).  
We speculate that the more metal-poor clusters may be
responsible for the larger velocity dispersion, and therefore should
have a more extended spatial distribution than the more metal-rich
ones.  This would support the possibility that some of these
metal-poor clusters belong to the intragalactic medium of the Fornax
cluster, as suggested by Muzzio (1987), Grillmair et al.\ (1994), and
West et al.\ (1995).

\acknowledgements { We would like to thank J. Rodriguez, V. Reyes, and
the whole NTT team for their expert and efficient support.  We thank
Jacco van Loon for taking acquisition exposures for us, and D. Geisler for
useful comments.  Work
performed at IGPP-LLNL is supported by the DOE under contract W7405-ENG-48.
MKP acknowledges the supported by the {\sc DFG} through the
Graduiertenkolleg `The Magellanic System and other dwarf galaxies', as
well as the hospitality of IGPP-LLNL during which part of this work was
done.  }

\clearpage

\begin{deluxetable}{rrr rl rrrrrrl}
\small
\footnotesize
\tablewidth{0pt}
\scriptsize
\tablecaption{Globular Cluster Candidates}
\tablehead{
\multicolumn{1}{c}{ID}&
\multicolumn{1}{c}{X}&
\multicolumn{1}{c}{Y}&
\multicolumn{1}{c}{RA(1950)}&
\multicolumn{1}{c}{DEC(1950)}&
\multicolumn{1}{c}{$V$}&
\multicolumn{1}{c}{$V\!-\!I$}&
\multicolumn{1}{c}{$v_{rad}$}&
\multicolumn{1}{c}{$\Delta_{Vr}$}&
\multicolumn{1}{c}{$\epsilon_{Vr}$}&
\multicolumn{1}{l}{Class} \\ [.2ex]
\colhead{}& \colhead{}&\colhead{}& \colhead{}&  \colhead{}& \colhead{}&
\colhead{}& \multicolumn{1}{c}{[km/s]} 
& \colhead{}& \colhead{}& \colhead{} }
\startdata
 201 &  529& $-$284&3 36 44.2 & -35 35 40 &   21.17& 1.24& 1061      & 20&135&glob\nl
 203 & 1532& $-$548&3 37 02.9 & -35 34 40 &   20.69& 1.08&  994      &115&73&glob\nl
 208 & 1107& $-$1382&3 36 55.0 & -35 31 31    &   20.91& 1.12& 1275      & 44& 91&glob\nl
 406 & $-$669& $-$199&3 36 21.9 &  -35 35 60   &   20.55& 1.11&  917 & 19&55&glob\nl
 407 & $-$1323& $-$450&3 36 09.8 &  -35 35 02  &   20.19& 1.01& 2107      & ---&159&glob\nl
 410 & $-$897& $-$791&3 36 17.7 &  -35 33 45   &   19.83& 1.27& 1190      & 13&94&glob\nl
 414 & $-$1013& $-$1077&3 36 15.5 &  -35 32 40 &   19.56& 1.09& 1565      & ---&105&glob\nl
 101 & 1129 & 1942 &3 36 55.4 &  -35 44 05 &  ---& ---& 1270 &      75&118&glob\nl
 103 & 745&1803& 3 36 48.3 &  -35 43 34 &          19.59& 0.63& 2738      & 49&398&blue glob?\nl
 104 & 1308&1744&3 36 58.7 &  -35 43 21 &          18.51& 1.48& 1459      & 17&52&(glob) dE\nl
 109 & 1315&1325&3 36 58.9 &  -35 41 46 &          21.24& 1.27& 1249      & 29&103&glob\nl
 113 & 337&1061&3 36 40.7 &  -35 40 46  &          21.15& 1.26& 1440      & 22&138&glob\nl
 119 & 1468&410&3 37 01.7 &  -35 38 18  &          21.16& 1.19& 1349 & 20&105&glob\nl
 122 & 1564&257&3 37 03.5 &  -35 37 43  &          20.77& 1.06& 1731      & 52&92&glob\nl
 123 & 1060&206&3 36 54.1 &  -35 37 32  &          20.93& 1.00& 1307      & 29&164&glob\nl
 124 & 461&126&3 36 43.0 &  -35 37 14   &          21.18& 1.25& 1142      & 16&189&glob\nl
 125 & 355&68&3 36 41.0 &  -35 37 00    &          21.02& 1.17& 1772      & 16&142&glob\nl
 126 & 553&29&3 36 44.7 &  -35 36 52    &          20.76& 1.22&  723      &132&207&glob\nl
 127 & 490& $-$57&3 36 43.5  & -35 36 32 &          21.06& 1.16& 1811      & 33&95&glob\nl
\enddata
\end {deluxetable}

\begin{deluxetable}{rrr rl rrrcrrl}
\small
\footnotesize
\tablewidth{0pt}
\scriptsize
\tablecaption{Background Galaxies and Foreground Stars}
\tablehead{
\multicolumn{1}{c}{ID}&
\multicolumn{1}{c}{X}&
\multicolumn{1}{c}{Y}&
\multicolumn{1}{c}{$V$}&
\multicolumn{1}{c}{$V\!-\!I$}&
\multicolumn{1}{c}{$v_{rad}$}&
\multicolumn{1}{c}{$\Delta_{Vr}$}&
\multicolumn{1}{c}{$\epsilon_{Vr}$}&
\multicolumn{1}{l}{Class} \\ [.2ex]
\colhead{}& \colhead{}&  \colhead{}& \colhead{}&
\colhead{}& \multicolumn{1}{c}{[km/s]}
& \colhead{}& \colhead{}& \colhead{} }
\startdata
 116 & 993&1545&   21.18& 1.10& 5400      & ---& ---&faint gal?\nl
 117 & 877&1504&   20.92& 1.22& 4560      & ---& ---&faint gal?\nl
 118 & 1091&1390&   21.17& 1.28& 3011      & ---& ---&faint gal?\nl
 120 & 1131&330&     21.65& 1.72& 2988     & ---& ---&faint gal\nl
 404 & $-$751& $-$73&     21.77& 1.89& 3105     & ---& ---&faint gal\nl
 412 & $-$478& $-$898&   20.70& 1.10& 4536\rlap{\,:}     & ---& ---&gal?\nl
 418 & $-$1084& $-$1678  &20.59& 0.83& 2849     & ---& ---&gal?\nl
\nl				      
 202 &   945& $-$452&  17.94& 0.83&    4      &39&24&star\nl
 204 &  1327& $-$716&  14.45& 0.58&   50      &30&14&ref star\nl
 205 &  1422& $-$816&  19.14& 1.78&  142      &69&51&star\nl
 207 &  828& $-$1105&  $<$14& 0.93&   15      &30&14&ref star\nl
 209 &  752& $-$1649&  18.22& 0.66&  219      &14&34&star\nl
 210 &  814& $-$1736&  18.31& 0.52&  $-$35      &42&44&star\nl
 211 &  932& $-$1809&  20.40& 1.89& $-$191      & ---&106&star\nl
 212 &  665& $-$1831&  19.58& 1.31&  201      &19&51&star\nl
 213 &  494& $-$1983&  17.70& 0.65&  120      &17&36&star\nl
 408 & $-$1328& $-$491 & 20.29& 1.69&   32      &5&70&star\nl
 409 & $-$1438& $-$647 &18.81& 1.33&  390      &2&43&star\nl
 411 & $-$1074& $-$815 & 19.81& 1.59&  133      &3&56&star\nl
 413 & $-$1209& $-$926 & 13.82& 0.80&  105      &12&8&ref star\nl
 419 & $-$964& $-$1702 & 18.82& 1.34&   88      &15&48&star\nl
 420 & $-$1227& $-$1790 &13.33& 0.36&   88      &12&8&ref star\nl
 415 & $-$1549& $-$1113 &19.50& 0.91&  144      &13&15&star\nl
 416 & $-$1086& $-$1540 &16.76& 0.97&  107      &21&20&star\nl
 417 &  $-$845& $-$1613 &17.57& 0.92&  120      &42&49&star\nl
 102 &    ---& ---&$<$14& ---&   $-$1      &25&22&ref star\nl
 106 &   993&1545&  19.24& 1.33&   72      &22&92&star\nl
 107 &   877&1504&  20.33& 0.82&   76      &4&88&star\nl
 108 &  1091&1390&  14.81& 1.33&  $-$24      &25&22&ref star\nl
 110 &  1093&1270  &19.80& 1.74&   91      &22&95&star\nl
 111 &   640&1208  &20.15& 0.92& $-$349      &43&284&star\nl
 112 &  1162&1160  &18.99& 2.38&   11      &27&100&star\nl
 114 &   863&796&  20.73& 1.73& $-$126      &50&89&star\nl
 115 &   821&746&   19.91& 0.78&  325      &39&89&star\nl
 121 &   545&292&   $<$14& ---&    1      &25&22&ref star\nl
\enddata
\end {deluxetable}

\begin{figure}
\psfig{figure=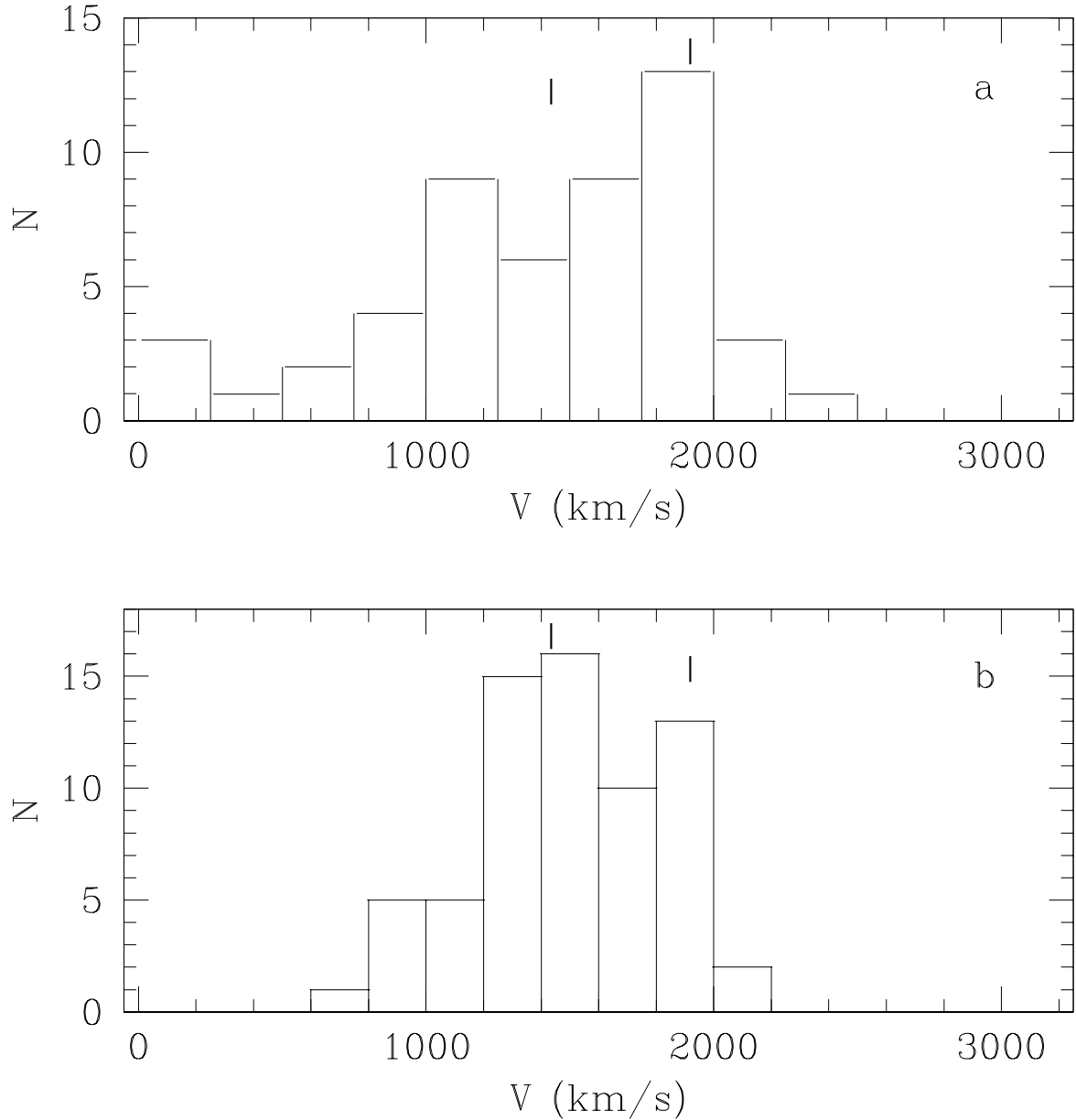,height=16cm,width=16cm
,bbllx=8mm,bblly=57mm,bburx=205mm,bbury=245mm}
\caption{
a) Radial velocity distribution for the globular clusters in NGC~1399,
from Grillmair et al.\ (1994).
b) Radial velocity distribution for the galaxies members of the Fornax
cluster, from Ferguson (1989).
The vertical tick marks in the figures, at $V=1294$ and $V=1944$,
indicate the radial velocities of the galaxies NGC~1399 and NGC~1404,
respectively.}
\end{figure}

\begin{figure}
\psfig{figure=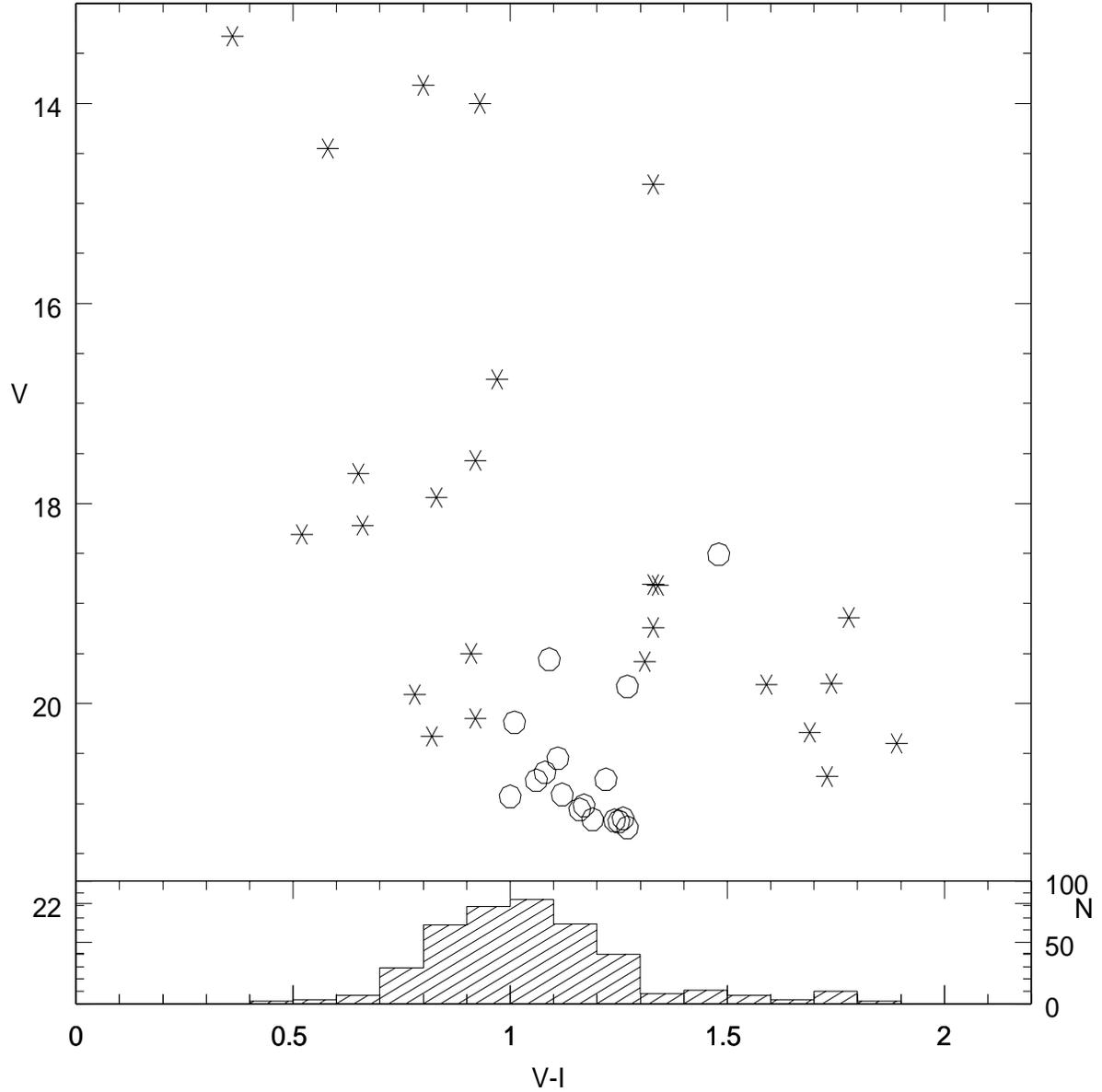,height=16cm,width=16cm
,bbllx=8mm,bblly=57mm,bburx=205mm,bbury=245mm}
\caption{Optical color-magnitude diagram for the non-saturated stars (asterisks)
and clusters (circles) of the sample. The brightest cluster candidate at
$V=18.5$, $V-I=1.5$ may actually be a dE galaxy (see text). The color distribution
of objects in the field of NGC~1399 is shown for comparison (from Kissler-Patig
et al. 1997). }
\end{figure}

\begin{figure}
\psfig{figure=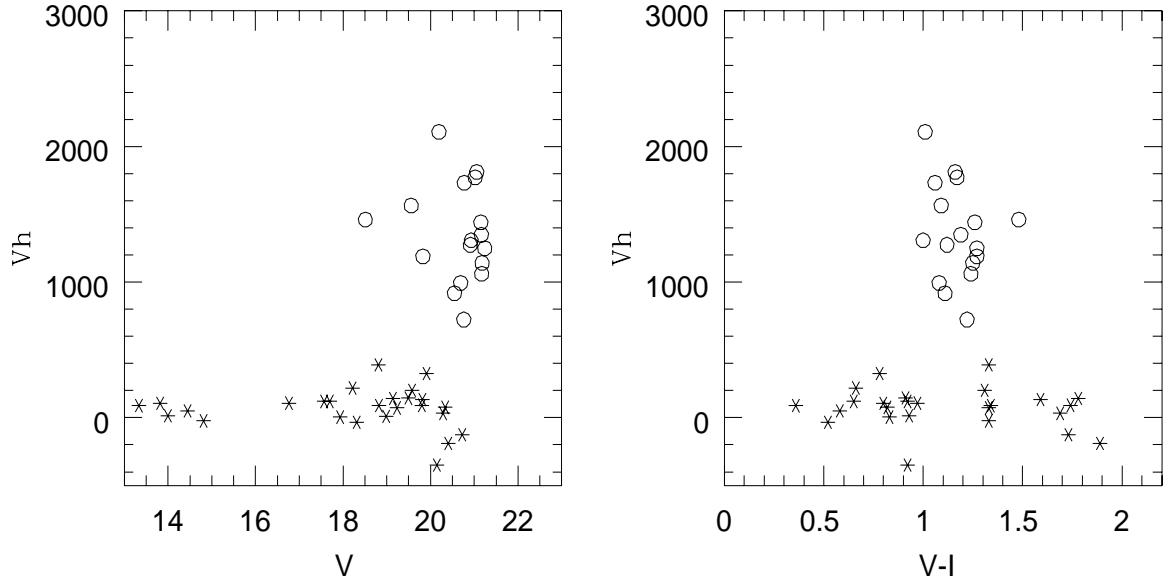,height=16cm,width=16cm
,bbllx=8mm,bblly=57mm,bburx=205mm,bbury=245mm}
\caption{ Dependence of velocity on $V$ magnitude (left panel) and $V-I$
color (right panel) for the stars (asterisks) 
and clusters (circles) of the sample.}
\end{figure}

\begin{figure}
\psfig{figure=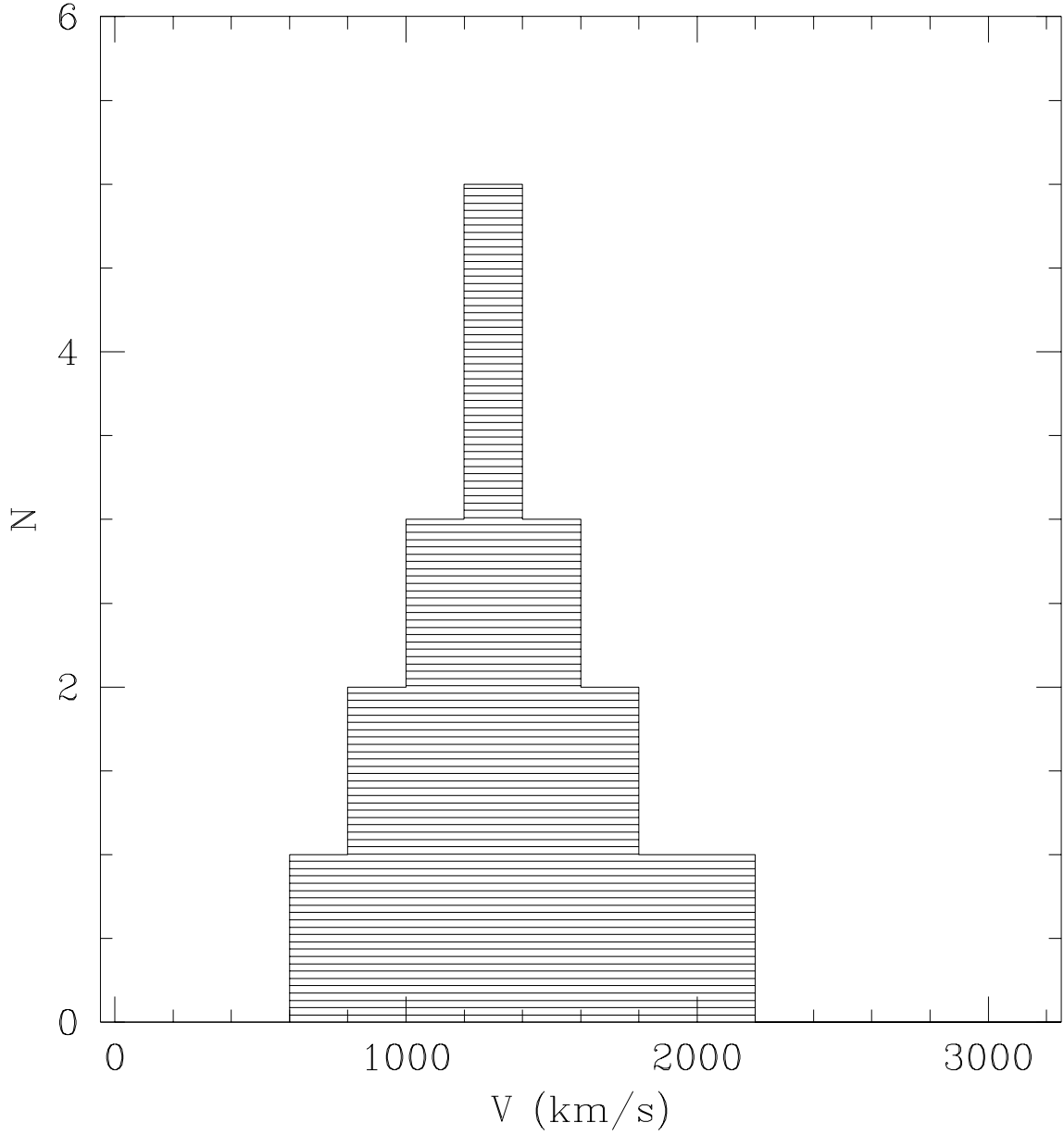,height=16cm,width=16cm
,bbllx=8mm,bblly=57mm,bburx=205mm,bbury=245mm}
\caption{ Velocity distribution for the 18 metal-rich globular
clusters in NGC~1399.}
\end{figure}

\begin{figure}
\psfig{figure=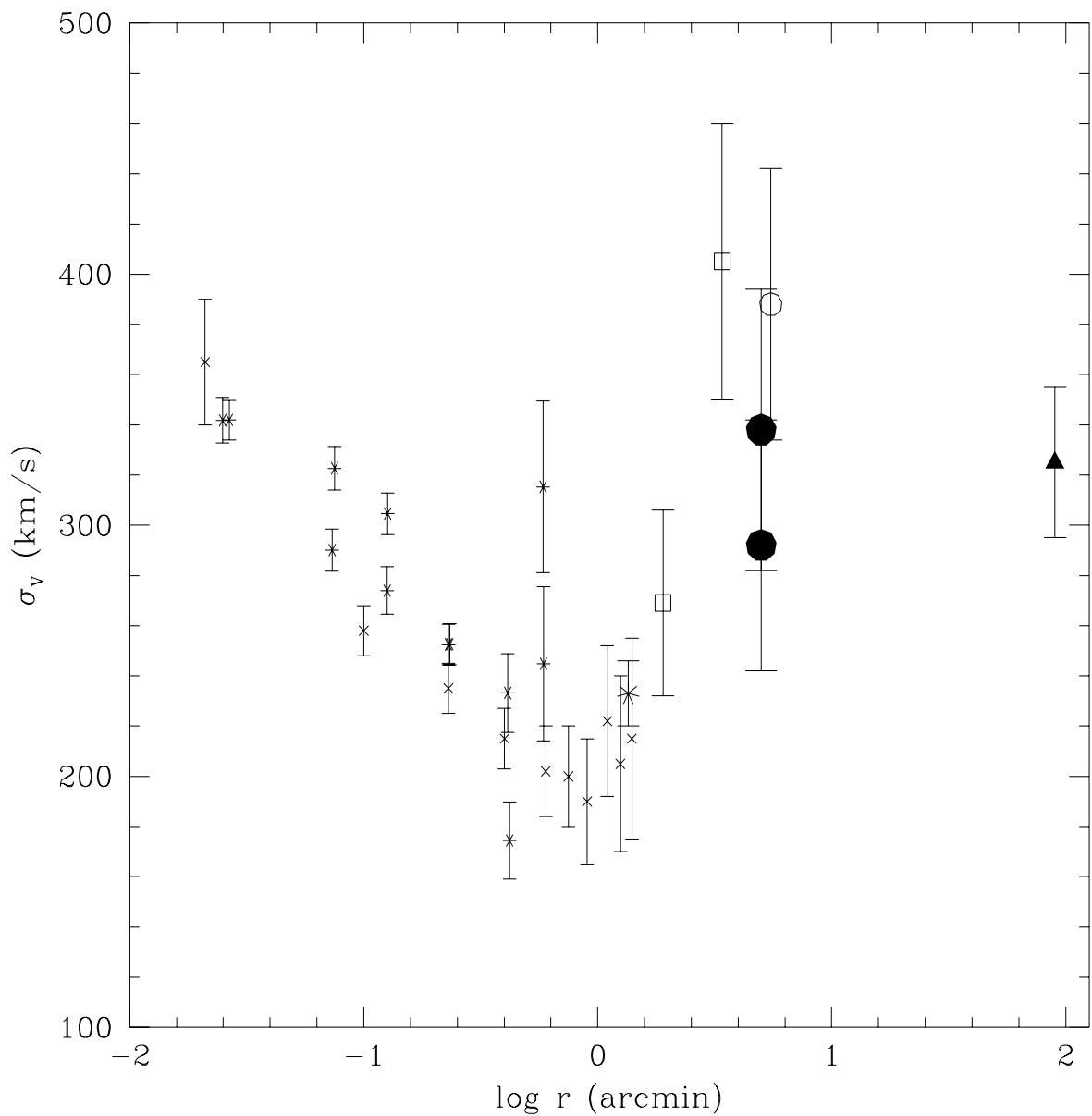,height=16cm,width=16cm
,bbllx=8mm,bblly=57mm,bburx=205mm,bbury=245mm}
\caption{ Radial dependence of the velocity dispersion in
NGC~1399. The stars and crosses are from the integrated stellar light
(Bicknell et al.\ 1989, Franx et al.\ 1989, Windsall \& Freeman 1994),
the open circle from the globular clusters of Grillmair et al.\
(1995), the squares from the planetary nebulae (Arnaboldi et al.\
1994), the triangle from galaxies of the Fornax cluster (Ferguson
1989), and the filled circles from metal-rich globular clusters from
this work (computed with and without the highest velocity cluster, see text).}
\end{figure}

\begin{figure}
\psfig{figure=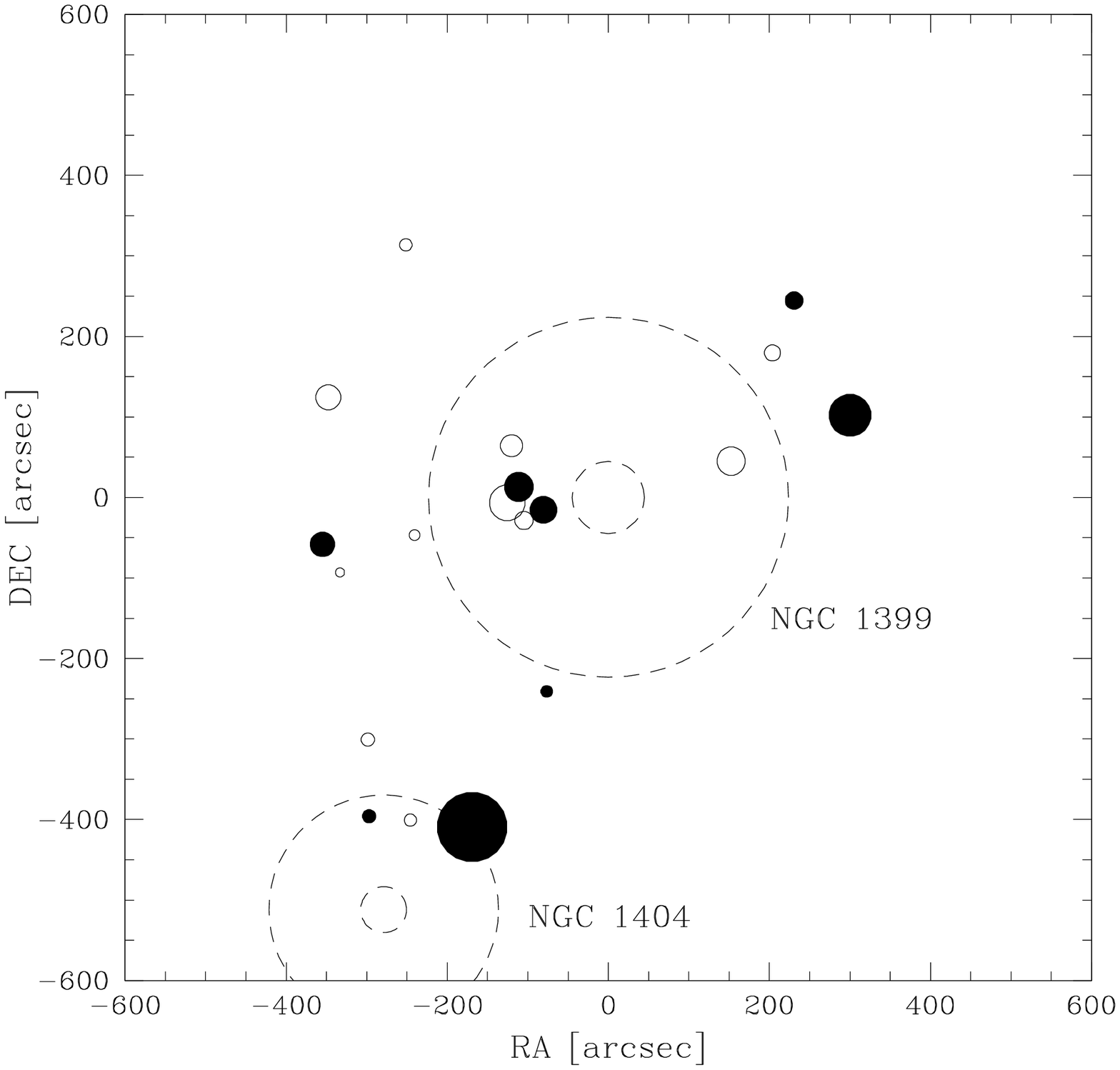,height=16cm,width=16cm
,bbllx=8mm,bblly=57mm,bburx=205mm,bbury=245mm}
\caption{ 
Spatial distribution of the confirmed globular cluster candidates with
good radial velocities. The RA and DEC axes are in pixels, with North up and
East to the left.   The dashed circles show the location 
of the galaxies NGC~1399 and NGC~1404, respectively. 
The open and solid circles represent objects with $V<1353$ and
$V>1353$ \kms, respectively, with their sizes scaling with the
velocity difference to the mean $V=1353$ \kms.
Note that the data of field 1 (towards 1404) have twice the S/N than 
the rest, and that on field 2 and 4 we have only 3 and 4 globular
clusters, respectively.
}
\end{figure}


\begin{references}
\reference{}{Armandroff, T., \& Zinn, R. 1988, AJ, 96, 92}
\reference{}{Arnaboldi, M., Freeman, K. C., Hui, X., Capaccioli, M., \& Ford, 
H. 1994, ESO Messenger, 76, 40}
\reference{}{Ashman, K. M., \& Bird, C. M. 1993, AJ, 106, 2281}
\reference{}{Ashman, K. M., Zepf, S. E. 1992, ApJ, 384, 50}
\reference{}{Baum, W.A., 1955 PASP, 67, 328}
\reference{}{Bassino, L. P., Muzzio, J. C., \& Rabolli, M.  1994 ApJ, 431, 634}
\reference{}{Beers, T. C., et al.\ 1991, AJ, 102, 1581}
\reference{}{Bica, E., \& Alloin, D. 1987, A\&A, 186, 49}
\reference{}{Bicknell, G. V., Carter, D., Killeen, N., \& Bruce, T. 1989, ApJ, 
336, 639}
\reference{}{Binney, J., \& Tremaine, S. 1997, Galactic Dynamics
(Princeton University Press)}
\reference{}{Bird, C. 1994, AJ, 107, 1637}
\reference{}{Blakeslee, J. P., \& Tonry, J. 1996, ApJ, 465, L19}
\reference{}{Bridges, T. J., Ashman, K., Zepf, S., et al.\ 1997, MNRAS, in 
press}
\reference{}{Bridges, T. J., Hanes, D. A., \& Harris, W. E. 1991, AJ, 101, 469}
\reference{}{Cohen, J. G., \& Ryzhov, A. 1997, AJ, in press (astro-ph/9704051)}
\reference{}{de Vaucouleurs, G. H., de Vaucouleurs, A., Corwin, H. G.,
Buta, R. J., Paturel, G., Fouqu\'e, P.\ 1991, ``Third Reference
Catalog of Bright Galaxies'' (New York: Springer)}
\reference{}{Durrell, P., Harris, W., Geisler, D. et al.\ 1996, AJ, 112, 972}
\reference{}{Elson, R. A. W., \& Santiago, B. X. 1996, MNRAS, 278, 617}
\reference{}{Ferguson, H. C. 1989, AJ, 98, 367}
\reference{}{Ferguson, H. C., \& Sandage, A. 1990, AJ, 100, 1}
\reference{}{Forbes, D. A., Brodie, J. P., \& Grillmair, C. J. 1997a, AJ, in 
press}
\reference{}{Forbes, D. A., Brodie, J. P., \& Huchra, J. 1997b, AJ, in press}
\reference{}{Forte, J. C., Strom, S. E., \& Strom, K. M. 1981, ApJ, 418, L55}
\reference{}{Franx, M., Illingworth, G., \& Heckman, T. 1989, ApJ, 344, 613}
\reference{}{Geisler, D., Lee, M. G., \& Kim, E. 1996, AJ, 111, 1529}
\reference{}{Geisler, D., \& Forte, J. C. 1990, ApJ, 350, L5}
\reference{}{Goudfrooij, P. et al.\ 1994, A\&AS, 104, 179}
\reference{}{Grillmair, C. J., Freeman, K., C.,  et al.\ 1994, ApJ, 422, L9}
\reference{}{Hanes, D. A., \& Harris, W. E. 1986, ApJ, 309, 564}
\reference{}{Hanes, D. A., \& Harris, W. E. 1986, ApJ, 309, 599}
\reference{}{Harris, W. E. 1991, ARA\&A, 29, 543}
\reference{}{Harris, W. E. 1995, in IAU Symp. 164, p. 85}
\reference{}{Harris, W. E., \& Hanes, D. A. 1987, AJ, 93, 1368}
\reference{}{Harris, W. E., Harris, G. L. H., \& Hesser, J. E. 1988, IAU Symp. 126, p. 215}
\reference{}{Heisler, J., Tremaine, S., \& Bahcall, J. N. 1985, ApJ, 298, 8}
\reference{}{Hill, J. M., Hintzen, P., Oegerle, W. R., Romanishin, W., Lesser, 
M. P., Eisenhamer, J. D., \& Batuski, D. J. 1988, ApJ, 332, L23}
\reference{}{Holtzman, J. A., et al. 1992, AJ, 103, 691}
\reference{}{Huchra, J., \& Brodie, J. 1987, AJ, 93, 779}
\reference{}{Hui, X., Ford, H. C., Freeman, K. C., \& Dopita, M. A. 1995, ApJ, 
449, 592}
\reference{}{Ikebe, Y., et al.\ 1996, Nature, 379, 427}
\reference{}{Kissler--Patig, M. 1997, A\&A, 319, 83}
\reference{}{Kissler--Patig, M., Kohle, S., M. Hilker, Richtler, T.,
et al.\ 1997, A\&A, 319, 470}
\reference{}{Kohle, S., Kissler-Patig, M., Hilker, M., Richtler,
T. R., Infante, L., Quintana, H.\ 1996, A\&A, 309, L39}
\reference{}{Minniti, D. 1995, AJ, 109, 1663}
\reference{}{Minniti, D., Alonso, M. V., Goudfrooij, P., Meylan, G., \& 
Jablonka, P. 1996, ApJ, 467, 221}
\reference{}{Mould, J., Oke, J. B., \& Nemec, J. M. 1987, AJ, 92, 53}
\reference{}{Mould, J., Oke, J. B., de Zeeuw, P. T., \& Nemec, J. M. 1990, AJ, 
99, 1823}
\reference{}{Muzzio, J. C. 1987, PASP, 99, 245}
\reference{}{N{\o}rgaard--Nielsen, H. U., Goudfrooij, P., J{\o}rgensen, H. E., 
\& Hansen, L. 1993, A\&A, 279, 61}
\reference{}{Oegerle, W. R., \& Hill, J. M. 1994, AJ, 107, 857}
\reference{}{Ostrov, P., Geisler, D., \& Forte, J. C. 1994, AJ, 105, 1762}
\reference{}{Poulain, P., 1988, A\&AS, 772, 215}
\reference{}{Racine, R., 1968, PASP 80, 326}
\reference{}{Richtler, T., Grebel, E. K., Domgorgen, H., Hilker, M., \& 
Kissler, M. 1992, A\&A, 264, 25}
\reference{}{Sharples, R. 1988, IAU Symp. 126, p. 245}
\reference{}{Sommer-Larsen, J. 1987, MNRAS, 227, 21P}
\reference{}{Strom, S. E., Forte, J. C., Harris, W. E., Strom, K. M., Wells, 
D. C., \& Smith, M. G. 1981, ApJ, 245, 416}
\reference{}{Theuns, T., \& Warren, S. J. 1997, MNRAS, 284, L11}
\reference{}{Tonry, J., \& Davis, M. 1979, AJ, 84, 1511}
\reference{}{Tully, R. B. 1988, Nearby Galaxies Catalog, Cambridge University 
Press}
\reference{}{van den Bergh, S. 1996, PASP, 108, 986}
\reference{}{de Vaucouleurs, G., et al.\ 1991, Revised Catalogue of
Galaxies Version 3.9 (RC3.9)}
\reference{}{Vazdekis, A., Casuso, E., Peletier, R. F., Beckman, J. E.\ 1996, 
ApJS, 106, 307}
\reference{}{Wagner, S., Richtler, T., \& Hopp, U. 1991, A\&A, 241, 399}
\reference{}{West, M., et al.\ 1996, ApJ, 453, L77}
\reference{}{Whitmore, B. C., Schweizer, F., Leitherer, C., Borne, K., \& 
Robert, C. 1993, AJ, 106, 1354}
\reference{}{Whitmore, B. C., \& Schweizer, F. 1995, AJ, 109, 960}
\reference{}{Whitmore, B. C., Sparks, W. B., Lucas, R. A., Macchetto, F. D., 
\& Biretta, J. A. 1995, ApJ, 454, L73}
\reference{}{Winsall, M., \& Freeman, K. C. 1993, A\&A, 268, 443}
\reference{}{Worthey, G. 1994, ApJS, 95, 107}
\reference{}{Zepf, S., \& Ashman, K. 1993, MNRAS, 264, 611}
\end{references}
\end{document}